\documentclass[12pt]{article}
\usepackage{a4wide}

\usepackage[english]{babel}
\usepackage{epsfig}
\usepackage{latexsym}
\usepackage{amsmath}
\usepackage{amssymb}
\usepackage{cite}

\def\ul(#1){\underline{#1}}

\usepackage{pslatex}

\usepackage{color,colordvi}

\begin{document}
\sloppy

\begin{titlepage}
\begin{flushright}
NIKHEF 11-008\\
\end{flushright}
\vspace{0.8cm}

\begin{center}
\Large
{\bf About a conjectured basis for Multiple Zeta Values}

\vspace*{1cm}
\large
J. Kuipers, J.A.M. Vermaseren
\\
\vspace{1.2cm}
\normalsize
{\it Nikhef \\
\vspace{0.1cm}
Science Park 105, 1098 XG Amsterdam, The Netherlands} \\
\vfill
\end{center}
\begin{abstract}
{\small
\noindent
We confirm a conjecture about the construction of basis elements for the 
multiple zeta values (MZVs) at weight 27 and weight 28. Both show as 
expected one element that is twofold extended. This is done with some 
lengthy computer algebra calculations using TFORM to determine explicit 
bases for the MZVs at these weights.
}  
\end{abstract}
\vfill
\end{titlepage}
%
%
\setcounter{equation}{0}


\section{Introduction}

Multiple Zeta Values (MZVs) have been around since Euler\cite{Euler} 
defined them. Recently interest in them was renewed by Zagier\cite{ZAG1} 
who formulated a number of conjectures about them. If one is to formulate a 
proof about irrationality of the MZVs it is of importance to know for each 
weight what constitutes a basis into which all elements with that weight 
can be expressed such that all coefficients are rational numbers. To derive 
such a basis, it is believed that there are two algebras that provide 
equations, and that there are no more equations than those\footnote{For 
definitions and notations we refer to ref.\cite{datamine}}. Solving the 
resulting sets of equations will determine a basis. In a recent 
publication\cite{datamine} these equations were solved to rather high 
values of the weight and depth and the number of basis elements was found 
in agreement with values conjectured by Broadhurst and Kreimer\cite{BK1}. 
In their conjecture Broadhurst and Kreimer made predictions about the 
distribution in weight and minimal depth of the basis elements and in 
addition considered the problem of pushdown\cite{Broadhurst:1} in which a 
basis element that over the MZVs can only be expressed with a given 
minimal depth can be expressed into Euler (alternating) sums of a lesser 
depth. They also gave a conjecture for the number of basis elements for 
which this occurs, but they considered only pushdowns over two units in 
depth.

In ref.\cite{datamine} a particular structure of potential basis elements 
was discovered. The conjectured number of basis elements when expressed for 
the Euler sums is equal to the number of Lyndon\cite{LYND} words of `depth' 
$D$ indices which add up to the weight and each having an odd integer value 
greater than one. In the MZV basis the elements that are pushed down in the 
Euler basis can be derived from a subset of the above set $L_W$ of Lyndon 
words of weight $W$ by subtracting one from the first two indices and 
adding two indices one at the end of the list as in
\begin{eqnarray}
	Z(m_1,m_2,m_3\cdots,m_D) & \rightarrow &
			Z(m_1-1,m_2-1,m_3\cdots,m_D,1,1).
\end{eqnarray}
For all values of the width and the depth that were accessible to computer 
evaluation such a basis could be constructed and the corresponding pushdown 
relations could be determined. This was done to weight 21. For larger 
values of the weight we could verify that such `one fold extended' bases 
could be constructed to weight 26 and partially (for limited depth) up to 
weight 30, but the pushdown relations were `out of range'. Yet there were 
already vague indications that this would not be enough and that for 
greater values of weight and depth so-called double extensions or worse 
might be needed. Here a $n$-fold extension is defined as that we take an 
element from the set $L_W$, lower the first $2n$ indices by one and add 
$2n$ indices with the value one at the end. A revised conjecture of the 
Broadhurst Kreimer generating functionals was constructed, resulting in 
table 18 in ref.\cite{datamine}. This table predicts that the first such 
twofold extended basis elements are to occur for $W=27, D=9$ and $W=28, 
D=8$. Unfortunately these runs were at the time exceeding our resources. 
The conjecture was formulated that these extensions and the pushdowns are 
coinciding and hence that for these values also double pushdowns should 
occur. To verify this explicitly by brute force is however at least six 
orders of magnitude beyond the current computer resources.

In the current publication we have made the FORM/TFORM\cite{FORM,TFORM} 
programs more efficient, both at the level of internal workings of FORM and 
the level of external FORM code. This has allowed us to run both cases and 
confirm the need for twofold extensions according to the conjecture.


\section{Improvements and running}

When we consider the original program as outlined in ref.\cite{datamine}, 
the first inefficiency we notice is that we use the stuffle relations to 
reduce all MZVs to those that are Lyndon words and then use the shuffle 
relations for the further reduction. As most of the computer time is in the 
substitution of eliminated elements in the master expression, we could save 
much time, if after using all the stuffle relations we store the result in 
a giant table (provided the computer has enough memory for this) and 
continue with only those MZVs of which the index field forms a Lyndon 
word. This should save roughly a factor $D$ (depth) in the number of MZVs 
in which we have to make these substitutions, although it goes at the cost 
of extra table lookups. At the same time it saves much disk space. It 
creates the need though to split the program in separate runs for each 
depth because originally we ran all depths in one program with the stuffle 
and shuffle relations run sequentially for each depth. This means that we 
should collect the results of the lower depths in terms of tables as well. 
In all it makes the structure of the program slightly more complicated, but 
it turned out to be worth it.

A second significant improvement is at the level of TFORM itself. The 
master expression is organized such that it has brackets and outside each 
bracket is the MZV, while inside is what it is currently equal to. When we 
send these terms inside the bracket to different workers and make 
substitutions on them, it means that cancellations between these terms take 
place only when the terms from the various workers are combined, i.e. very 
late in the sorting. We measured for the run at $W=28$ that the combined 
sort files of the workers took more than 300 Gbytes while the output file 
took only just over 100 Gbytes. This also made it enter a new level of 
complexity in which the sorting needed an extra stage for which another 
300+ Gbytes were needed. In total the program needed almost 800 Gbytes of 
disk space. By giving TFORM a mode in which it distributes complete 
brackets over the workers the cancellations took all place inside each 
worker rather than when the results of the workers were combined. The 
result is that fewer compare operations are needed and that the combined 
sort files of the workers take hardly more space than the full output. Also 
the extra sorting stage was no longer needed. This meant that use of space 
for the above case would have shrunk to slightly over 200 Gbytes. Together 
with the first improvement we managed to limit disk space to less than 42 
Gbytes. Because this is an improvement in TFORM, all future programs 
benefit from it automatically.

A third improvement found its origin in the observation that the stuffle 
relations at a given depth mix only MZVs that have the same indices, 
albeit in a different order. This means that we can work though those 
`families' one by one without having the complete master expression under 
consideration. Each time a family has been completed, the results can be 
placed in the tables. Although the part with the stuffle relations uses 
only a fraction of the CPU time, it was still possible to save more than a 
day running time by using this method. The added complexity was only 
moderate.

Another improvement which we did not use in the end is to consider the 
stuffle operations and use them to transform each MZV `closer' to a linear 
combination of Lyndon words. This must be programmed very carefully. In the 
end we managed to get a version which was faster than the above way of 
treating the stuffle relations, but because the profit was not very big and 
the program was more complicated, we did not use it in the final runs.

The first run we made was with the original program that was also used in 
ref.\cite{datamine}. This was before all the above improvements were made. 
At the computer in Karlsruhe\footnote{Both the Karlsruhe and the DESY 
computers we used have 64-bits processors running at 3.2 GHz.} this run 
took 69 days. Then, when we had made the first two improvements we reran 
the program and the run for $D=8$ took 28 days (the combined runs for the 
lower depths took 40+ hours). As one can see, we did not gain a factor 8 as 
hoped for when improvement one was made. The extra table substitutions seem 
to be part of the problem. In addition the substitutions in the Lyndon 
words seem to be more complex. This is confirmed by the fact that in the 
first run, the largest size of the master expression was more than 100 
Gbytes while in the second run it was slightly under 21 Gbytes which is not 
a factor 8 as one would hope for initially.

The problem with the run at $W=27$, $D=9$ was to hold all tables in memory. 
The Karlsruhe computer, with 32 Gbytes of CPU memory was not up to this 
task. Fortunately at DESY there was a computer with 8 processors and 96 
Gbytes of memory. The completely optimized program was started there and 
finished 85 days later. The largest need for disk space was twice 35 
Gbytes. During the run the program processed more than $3.1\times 10^{13}$ 
terms. On average this is more than 500000 terms per core per second and, 
considering the average size of the terms (at least 68 bytes), the sort 
routines had to process more than 2 Petabytes of information.

We tested the results of the $W=27$ run by generating all relevant shuffle 
relations again and directly substituting the tables generated by the 
previous runs. All relations collapsed to zero.


\section{Results}

Having run the $W=27$ and $W=28$ programs we can now construct the complete 
bases for these cases in a `minimal depth' representation.

\begin{eqnarray}
	  P_{27} & = &
		\!H_{27},
		\!H_{11,7,9},
		\!H_{13,11,3},
		\!H_{15,3,9},
		\!H_{15,5,7},
		\!H_{15,7,5},
		\!H_{15,9,3},
		\!H_{17,5,5},
		\!H_{17,7,3},
			\nonumber \\ &&
		\!H_{19,3,5},
		\!H_{19,5,3},
		\!H_{21,3,3},
		\!H_{7,5,5,7,3},
		\!H_{7,5,7,3,5},
		\!H_{7,7,3,7,3},
		\!H_{7,7,7,3,3},
			\nonumber \\ &&
		\!H_{9,3,9,3,3},
		\!H_{9,5,3,5,5},
		\!H_{9,5,3,7,3},
		\!H_{9,5,5,3,5},
		\!H_{9,5,5,5,3},
		\!H_{9,5,7,3,3},
		\!H_{9,7,3,3,5},
			\nonumber \\ &&
		\!H_{9,7,3,5,3},
		\!H_{9,7,5,3,3},
		\!H_{9,9,3,3,3},
		\!H_{11,3,3,3,7},
		\!H_{11,3,3,5,5},
		\!H_{11,3,3,7,3},
		\!H_{11,3,5,3,5},
			\nonumber \\ &&
		\!H_{11,3,5,5,3},
		\!H_{11,3,7,3,3},
		\!H_{11,5,3,3,5},
		\!H_{11,5,3,5,3},
		\!H_{11,5,5,3,3},
		\!H_{11,7,3,3,3},
			\nonumber \\ &&
		\!H_{13,3,3,3,5},
		\!H_{13,3,3,5,3},
		\!H_{13,3,5,3,3},
		\!H_{13,5,3,3,3},
		\!H_{15,3,3,3,3},
		\!H_{10,8,7,1,1},
			\nonumber \\ &&
		\!H_{10,10,5,1,1},
		\!H_{12,2,11,1,1},
		\!H_{12,4,9,1,1},
		\!H_{12,6,7,1,1},
		\!H_{12,8,5,1,1},
		\!H_{16,2,7,1,1},
			\nonumber \\ &&
		\!H_{5,3,5,3,5,3,3},
		\!H_{5,5,3,3,3,5,3},
		\!H_{5,5,3,3,5,3,3},
		\!H_{5,5,3,5,3,3,3},
		\!H_{5,5,5,3,3,3,3},
			\nonumber \\ &&
		\!H_{7,3,3,3,3,3,5},
		\!H_{7,3,3,3,3,5,3},
		\!H_{7,3,3,3,5,3,3},
		\!H_{7,3,3,5,3,3,3},
		\!H_{7,3,5,3,3,3,3},
			\nonumber \\ &&
		\!H_{9,3,3,3,3,3,3},
		\!H_{6,4,5,5,5,1,1},
		\!H_{6,6,3,5,5,1,1},
		\!H_{6,6,5,3,5,1,1},
		\!H_{6,6,5,5,3,1,1},
			\nonumber \\ &&
		\!H_{8,2,3,5,7,1,1},
		\!H_{8,2,3,7,5,1,1},
		\!H_{8,2,5,3,7,1,1},
		\!H_{8,2,5,5,5,1,1},
		\!H_{8,2,5,7,3,1,1},
			\nonumber \\ &&
		\!H_{8,2,7,3,5,1,1},
		\!H_{8,2,7,5,3,1,1},
		\!H_{8,4,3,3,7,1,1},
		\!H_{6,4,3,3,3,3,3,1,1},
		\!H_{6,4,6,4,3,1,1,1,1}
\end{eqnarray}
\begin{eqnarray}
	  P_{28} & = &
		\!H_{19,9},           
		\!H_{21,7},           
		\!H_{23,5},           
		\!H_{25,3},           
		\!H_{9,5,7,7},        
		\!H_{9,7,9,3},        
		\!H_{11,7,7,3},       
		\!H_{11,9,3,5},       
		\!H_{11,9,5,3},       
			\nonumber \\ &&
		\!H_{11,11,3,3},      
		\!H_{13,3,3,9},       
		\!H_{13,3,5,7},       
		\!H_{13,3,7,5},       
		\!H_{13,3,9,3},       
		\!H_{13,5,3,7},       
		\!H_{13,5,5,5},       
			\nonumber \\ &&
		\!H_{13,5,7,3},       
		\!H_{13,7,3,5},       
		\!H_{13,7,5,3},       
		\!H_{13,9,3,3},       
		\!H_{15,3,5,5},       
		\!H_{15,3,7,3},       
		\!H_{15,5,3,5},       
			\nonumber \\ &&
		\!H_{15,5,5,3},       
		\!H_{15,7,3,3},       
		\!H_{17,3,3,5},       
		\!H_{17,3,5,3},       
		\!H_{17,5,3,3},       
		\!H_{19,3,3,3},       
		\!H_{14,12,1,1},      
			\nonumber \\ &&
		\!H_{16,10,1,1},      
		\!H_{5,5,5,5,5,3},    
		\!H_{7,3,5,7,3,3},    
		\!H_{7,3,7,3,3,5},    
		\!H_{7,3,7,3,5,3},    
		\!H_{7,5,3,3,7,3},    
			\nonumber \\ &&
		\!H_{7,5,3,5,5,3},    
		\!H_{7,5,3,7,3,3},    
		\!H_{7,5,5,3,5,3},    
		\!H_{7,5,7,3,3,3},    
		\!H_{7,7,3,3,5,3},    
		\!H_{7,7,5,3,3,3},    
			\nonumber \\ &&
		\!H_{9,3,3,3,5,5},    
		\!H_{9,3,3,3,7,3},    
		\!H_{9,3,3,5,3,5},    
		\!H_{9,3,3,5,5,3},    
		\!H_{9,3,3,7,3,3},    
		\!H_{9,3,5,3,3,5},    
			\nonumber \\ &&
		\!H_{9,3,5,3,5,3},    
		\!H_{9,3,5,5,3,3},    
		\!H_{9,3,7,3,3,3},    
		\!H_{9,5,3,3,3,5},    
		\!H_{9,5,3,3,5,3},    
		\!H_{9,5,3,5,3,3},    
			\nonumber \\ &&
		\!H_{9,5,5,3,3,3},    
		\!H_{9,7,3,3,3,3},    
		\!H_{11,3,3,3,3,5},   
		\!H_{11,3,3,3,5,3},   
		\!H_{11,3,3,5,3,3},   
		\!H_{11,3,5,3,3,3},   
			\nonumber \\ &&
		\!H_{11,5,3,3,3,3},   
		\!H_{13,3,3,3,3,3},   
		\!H_{8,6,5,7,1,1},    
		\!H_{8,8,3,7,1,1},    
		\!H_{8,8,5,5,1,1},    
		\!H_{8,8,7,3,1,1},    
			\nonumber \\ &&
		\!H_{10,2,5,9,1,1},   
		\!H_{10,2,7,7,1,1},   
		\!H_{10,2,9,5,1,1},   
		\!H_{10,4,3,9,1,1},   
		\!H_{10,4,5,7,1,1},   
		\!H_{10,4,7,5,1,1},   
			\nonumber \\ &&
		\!H_{10,4,9,3,1,1},   
		\!H_{10,6,3,7,1,1},   
		\!H_{10,6,5,5,1,1},   
		\!H_{14,2,3,7,1,1},   
		\!H_{5,3,3,5,3,3,3,3},
			\nonumber \\ &&
		\!H_{5,3,5,3,3,3,3,3},
		\!H_{5,5,3,3,3,3,3,3},
		\!H_{7,3,3,3,3,3,3,3},
		\!H_{6,2,3,5,5,5,1,1},
		\!H_{6,2,5,3,5,5,1,1},
			\nonumber \\ &&
		\!H_{6,2,5,5,3,5,1,1},
		\!H_{6,2,5,5,5,3,1,1},
		\!H_{6,4,3,3,5,5,1,1},
		\!H_{6,4,3,5,3,5,1,1},
		\!H_{6,4,5,3,3,5,1,1},
			\nonumber \\ &&
		\!H_{6,4,5,5,3,3,1,1},
		\!H_{6,6,3,3,3,5,1,1},
		\!H_{6,6,3,5,3,3,1,1},
		\!H_{8,2,3,3,3,7,1,1},
		\!H_{8,6,6,4,1,1,1,1}
\end{eqnarray}
The above bases are completely in agreement with the predictions of table 
18 in ref.\cite{datamine}.
They have the property that if we take the elements with n trailing 
ones and add those to the first n elements as in
\begin{eqnarray}
	H_{6,4,6,4,3,1,1,1,1} & \rightarrow & H_{7,5,7,5,3}
\end{eqnarray}
we obtain the complete sets $L_{27}$ and $L_{28}$ respectively. These are 
the sets of all Lyndon words of odd integers greater than one that add up 
to $27$ and $28$ respectively.

The above bases are minimal in depth. This means that it is impossible to 
find a basis in which the sum of the depth of all elements is less.


\section{Discussion}

Considering that the origin of table 18 in ref.\cite{datamine} is related 
to pushdowns which could be tested up to weight 21, the fact that the same 
table describes the structure of the basis elements of the MZVs and was 
capable of predicting the twofold extended elements at the weights 27 and 28 
is a very strong indication that both phenomena are related as was already 
conjectured in ref.\cite{datamine}. Unfortunately, due to the current 
status of computer technology, it is very difficult to use our programs to 
obtain more results soon. Now is the time for more theoretical 
contributions, c.f. ref.\cite{Brown:2011ik}.

This work has been supported by the Dutch Foundation for Fundamental 
Research of Matter (FOM). We would like to thank the university of 
Karlsruhe and DESY for the use of their computers and J. Bl\"umlein for 
reading the manuscript.



\begin{thebibliography}{99}
%
\bibitem{Euler} 
L. Euler, {\it Meditationes circa singulare serium genus},
Novi Comm. Acad. Sci. Petropol. {\bf 20} (1775) 140--186, reprinted in {\sf Opera Omnia}
ser I vol. 15, (B.G. Teubner, Berlin, 1927), 217--267.
%
\bibitem{ZAG1} 
D. Zagier, {\it Values of zeta functions and their applications}, in~: First European Congress 
of Mathematics, Vol. II, (Paris, 1992), Progr. Math., {\bf 120}, (Birkh\"auser, Basel--Boston, 1994), 
pp.~497--512.
%
%
\bibitem{datamine} 
  J.~Bl\"umlein, D.~J.~Broadhurst and J.~A.~M.~Vermaseren,
  {\it The Multiple Zeta Value Data Mine},
  Comput. Phys. Commun. {\bf 181} (2010) 582-625,
  {\tt arXiv:math-ph/09072557.}
%
%
\bibitem{BK1}
  D.~J.~Broadhurst and D.~Kreimer,
  {\it Association of multiple zeta values with positive knots via Feynman
  diagrams up to 9 loops},
  Phys.\ Lett.\  B {\bf 393} (1997) 403--412, 
  {\tt [arXiv:hep-th/9609128]}.
%
\bibitem{Broadhurst:1} 
  D.~J.~Broadhurst,
  {\it On the enumeration of irreducible k-fold Euler sums and their roles in knot
  theory and field theory},
  {\tt arXiv:hep-th/9604128}.
%
\bibitem{LYND}
R.C. Lyndon, 
{\it On Burnsides problem}, 
Trans. Amer. Math. Soc. {\bf 77} (1954) 202--215;
{\it On Burnsides problem II}, 
Trans. Amer. Math. Soc. {\bf 78} (1955) 329--332;\\
C. Reutenauer, {\it Free Algebras}, (Calendron Press, Oxford, 1993).
%
\bibitem{FORM}
  J.~A.~M.~Vermaseren,
  {\it New features of FORM}, 
  {\tt arXiv:math-ph/0010025.}
%
\bibitem{TFORM} 
  M.~Tentyukov and J.~A.~M.~Vermaseren,
  {\it The multithreaded version of FORM},
  Computer.Phys.Commun 181(2010 1419-1427
  {\tt arXiv:hep-ph/0702279.}

\bibitem{Brown:2011ik}
  F.~Brown,
  {\it On the decomposition of motivic multiple zeta values},\hfill \\
  {\tt arXiv:1102.1310 [math.NT]}.


\end{thebibliography}
\end{document}